\documentclass[aps,prb,twocolumn,showpacs,groupedaddress]{revtex4}
\usepackage{graphicx}  
\usepackage{ulem}
\usepackage{dcolumn}   
\usepackage{bm}        
\usepackage{amssymb}   
\usepackage{epstopdf}
\usepackage{color}
\usepackage{dcolumn}
\usepackage{bm}
\def\gapx{\lower 2pt \hbox{$\buildrel>\over{\scriptstyle{\sim}}$\ }}
\def\lapx{\lower 2pt \hbox{$\buildrel<\over{\scriptstyle{\sim}}$\ }}

\RequirePackage[colorlinks,linkcolor=black,citecolor=blue,hyperindex]{hyperref}

\begin{document}
\title{Pairing and chiral spin density wave instabilities on the honeycomb lattice: a  comparative quantum Monte Carlo study}

\author{Tao Ying$^{1,2}$ and Stefan Wessel$^{1}$}
\affiliation{$^1$Institut f\"ur Theoretische Festk\"orperphysik, JARA-FIT and JARA-HPC, RWTH Aachen University, 52056 Aachen, Germany}
\affiliation{$^2$Department of Physics, Harbin Institute of Technology, 150001 Harbin, China}

\date{\today}

\begin{abstract}
Using finite-temperature determinantal quantum Monte Carlo calculations, we re-examine the pairing susceptibilities in the Hubbard model on the honeycomb lattice, focusing on  doping levels onto and away from the van Hove singularity (VHS) filling. For this purpose,   electronic densities of $0.75$ (at the hole-doping VHS) and $0.4$ (well below the VHS) are considered in detail, where due to a severe sign problem at strong coupling strengths, we focus on the weak interaction region of the Hubbard model Hamiltonian. From analyzing the temperature dependence of  pairing  susceptibilities in various symmetry channels, we find the singlet $d$+$id$-wave to be the dominant pairing channel both at and away from the VHS filling. We furthermore investigate the electronic susceptibility to a specific chiral spin density wave (SDW) order, which we find to be similarly relevant at the VHS, while it extenuates upon doping away from the VHS filling.
\end{abstract}

\maketitle
\section{Introduction}
In recent years, graphene~\cite{Novoselov04,Neto09,Kotov12} has attracted a lot of attentions, due to its unusual
electronic properties.  At charge neutrality, corresponding to a half-filled lattice in the Hubbard model description of  graphene's $\pi$-electron system, a vanishing density of states at the Fermi level (the Dirac points) renders
a semi-metallic state stable against instabilities from electron-electron interactions, even in the intermediate coupling regime~\cite{Sorella92,Paiva05,Herbut06,Herbut09}.
In contrast, upon doping well away from the  Dirac points through chemical doping~\cite{McChesney10} or electrical gating~\cite{Efetov10},
correlation effects are expected to no longer be limited to the strong interaction regime. Indeed, various possible  phases, such as superconducting instabilities, magnetism or charge/spin density waves have been considered to emerge in doped graphene:
%
%
Several theoretical studies focused on superconducting states of correlated electrons on the honeycomb lattice of graphene, mainly within a local Hubbard model description~\cite{Black-Schaffer14}. Based on mean field theory, Black-Schaffer {\it et al.}~\cite{Black-Schaffer07},
suggest that graphene  may become  a $d+id$-wave superconductor over a wide range of doping, while Uchoa {\it et al.}\cite{Uchoa07} suggest extended $s$-wave and $p+ip$-wave pairing states. Functional renormalization group (fRG) theory calculations proposed $f$-wave and $d+id$-wave instabilites~\cite{Honerkamp08}, and variational Monte Carlo~\cite{Pathak10,Watanabe13} and auxiliary-field quantum Monte Carlo study\cite{Ma11} both support $d+id$-wave pairing, while a variational cluster approximation and a cellular dynamical mean-field theory study~\cite{Faye15} suggests a $p+ip$ pairing symmetry. In general, this problem is thus far from having reached a conclusion.
An even more peculiar  condition is obtained upon doping the electronic system onto the van Hove singularity (VHS), where the non-interacting extended Fermi surface exhibits perfect nesting.
As a consequence, the pairing mechanism may be different from the one at more generic doping levels~\cite{Gonzlez08,Lamas09},
and furthermore the electronic system might even host other types of orders, such as a Pomeranchuk instability~\cite{Valenzuela08} or
a chiral spin density wave (SDW) order~\cite{Li12}. Different scenarios have indeed been proposed: A renormalization group study  finds $d+id$ pairing at the VHS filling in the weak coupling limit~\cite{Nandkishore12}.
Using an fRG approach, Wang {\it et al.} obtained  a chiral SDW
in the intermediate interaction region at the VHS filling, while  $d+id$ pairing was obtained away from the VHS\cite{wang12}.
Another fRG study reports possible $d+id$ or SDW  instabilities in the intermediate interaction region at the VHS, and $d+id$ or $f$-wave pairing away from the VHS~\cite{Kiesel12}.
More recently a dynamic cluster approximation study suggests that the $d+id$-wave pairing state dominates in the weak-coupling regime, while for  stronger interactions, a $p+ip$-wave state strongly competes with the $d+id$-wave state~\cite{Xu16}. However, in this study, SDW instabilities have not been considered.
This states of affairs motivates us  to examine this problem using finite-temperature determinantal quantum Monte Carlo (FT-DQMC), an essentially un-biased numerical algorithm. The rest of this paper is organized as follows: In Sec. II, we introduce the model that we consider and outline the FT-DQMC approach. Then we analyze in Sec. III various pairing channels of superconducting instabilities, while in Sec. IV, we consider the chiral SDW instability and contrast its behavior to  other magnetic ordering channels. Finally, we summarize our results in Sec. V.

\section{Model and Method}

In this paper, we examine the effective pairing susceptibility for various different pairing channels , and identify  the dominant pairing channel for doping levels onto and away from the VHS. Moreover, we  also consider  the chiral SDW  instability  that was  proposed by Li~\cite{Li12}, and examine,  to what extend this chiral SDW instability  effects  the behavior at the VHS filling and upon doping  away from the VHS point. For this analysis, we consider
the Hubbard model on the honeycomb lattice to describe the doped  graphene system. This model is given in terms of the  Hamiltonian
\begin{eqnarray}
\label{Hamiltonian}
{H}&=&-t\sum_{\langle { \bf i , \, j} \rangle \, \sigma}
(c_{{\bf i}\sigma}^\dagger c_{{\bf j}\sigma}^{\phantom{\dagger}} +
c_{{\bf j}\sigma}^\dagger c_{{\bf i}\sigma}^{\phantom{\dagger}} )
+ U \sum_{\bf i} n_{{\bf   i}\uparrow} n_{{\bf i} \downarrow}
\nonumber
\\
&&-\mu \sum_{{\bf i}}
(n_{{\bf i}\uparrow} + n_{{\bf i}\downarrow}),
\label{hamiltonian}
\end{eqnarray}
where $t$ is the fermion hopping amplitude between
nearest neighbor sites on the honeycomb lattice (here, ${\bf i}$ and ${\bf j}$ denote lattice vectors), $U$ denotes
an onsite repulsion,  and $\mu$ the chemical potential that allows to  tune the
electron density, denoted $\rho$ in the following. We work in units of $t=1$ in the following.

\begin{figure}[t]
\centering
\includegraphics[width=0.7\columnwidth,angle=0,clip]{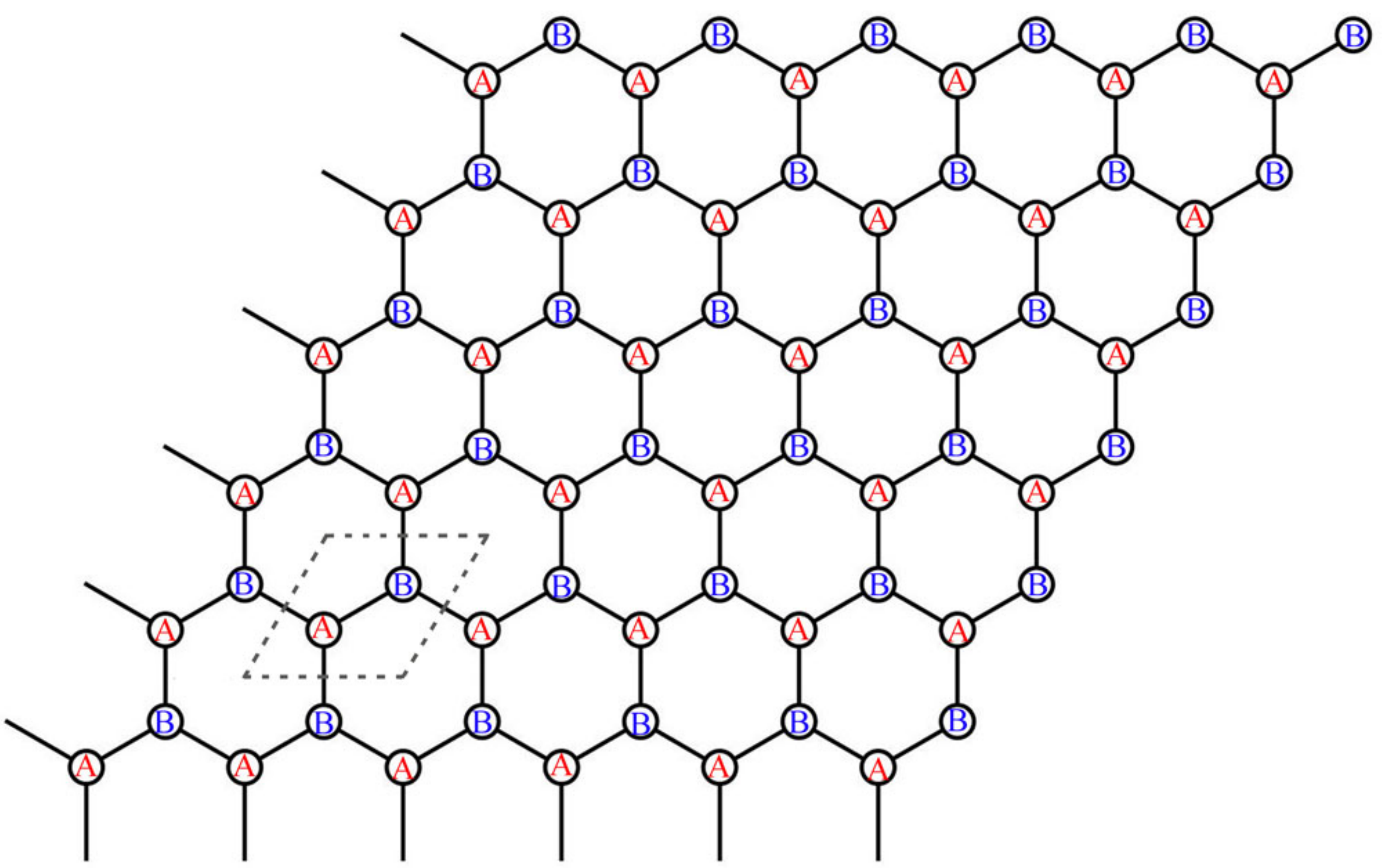}
\caption{(Color online)  Rhombic honeycomb lattice geometry for $L=6$, with $N_s=72$ sites. Dashed lines enclose the two-site unit cells, and the bipartite
sublattice structure is indicated by  site-centered letters $A$ and $B$.
 \label{fig:honeycomb} }
\end{figure}

The numerical algorithm used in this paper is the finite-temperature determinantal quantum Monte Carlo (FT-DQMC) method~\cite{Blankenbecler81,white89}.
We consider finite rhombic clusters of the bipartite honeycomb lattice with periodic boundary conditions and
with $N_s=L \times L \times 2$ lattice sites, mainly for $L=6$ and $L=12$ in order to ensure that both the $K$ (Dirac) and the $M$ points of the hexagonal Brillouin zone are included in the discrete lattice momentum space.
Close to  the VHS filling, we also consider other even linear system sizes such as $L=10$ and $L=14$ (for $L$ even, the $M$ points are included in the discrete lattice momentum space).
The finite lattice geometry for $L=6$ in real space is shown in Fig.~\ref{fig:honeycomb}.
The simulations were performed  at finite temperatures, and  we then analyzed the observed tendencies upon lowering the temperature.
In the following, we are mainly interested in  the doping level of the VHS, where the electron density is $\rho=0.75$ or $\rho=1.25$.
Due to particle-hole symmetry, we considered the case of $\rho=0.75$ explicitly.

\begin{figure}[t]
\centering
\includegraphics[width=\columnwidth,angle=0,clip]{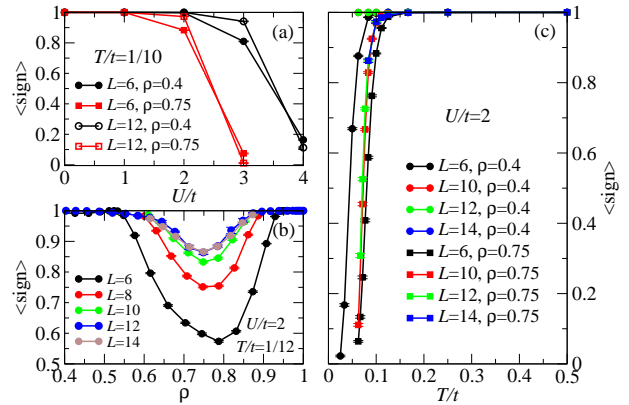}
\caption{(Color online) The average FT-DQMC sign, $\langle \mathrm{sign}\rangle$, for different lattice sizes as a function of (a) interaction strength $U$, (b) density $\rho$ and (c) temperature $T$. \label{sign_all} }
\end{figure}

However, upon doping beyond  half-filling, the FT-DQMC method suffers from a severe sign problem, which  worsens upon lowering the temperature and increasing the  interaction strength~\cite{Iglovikov15}.
To  quantify the sign-problem of the FT-DQMC in the relevant parameter regime, we show
in Fig.~\ref{sign_all} the dependence of the average sign, $\langle \mathrm{sign}\rangle$, on the interaction strength  $U$, the density $\rho$, and the  temperature $T$ for different  lattice sizes.
As Monte Carlo errors decrease with the square root of the number of independent samples, it is necessary to run a simulation code $100$ times longer
to compensate, for example, for an average sign of $0.1$.
Figure~\ref{sign_all} (a) shows that at the VHS filling of $\rho=0.75$, the average sign rapidly drops to values below 0.1 beyond $U=2t$ at the considered temperature of $T/t=1/10$. Furthermore, a dip in  $\langle \mathrm{sign}\rangle$ at the VHS filling of $\rho=0.75$ is  seen in the density dependence of  $\langle \mathrm{sign}\rangle$  in  Fig.~\ref{sign_all} (b) for $U/t=2$. For the considered  temperature of $T/t=1/12$, this dip is more pronounced for the smaller system sizes, while the average sign appears to converge upon increasing $L$ at this fixed temperature  to a still conveniently large value.
However, as seen from Fig.~\ref{sign_all} (c), the average sign shows a  rapid drop  with decreasing temperature also for $U/t=2$, which restricts us from accessing true ground-state properties on large systems near the VHS filling.

Hence, depending on the doping level, and in particular near the VHS filling, we restricted our investigation to the weak to intermediate interaction regime, in order to still access low temperatures that allow us to identify the onset of  divergences in  the pairing or magnetic susceptibilites.
Furthermore, in order to compare the results for the VHS filling with those at more generic fillings,
we also  performed further simulations at $\rho=0.4$, i.e.,  a doping level far below the VHS filling, and where the sign problem
is less severe, and  we can extend a bit further towards the stronger interaction regime.

\section{Pairing Correlations}

\begin{figure}[t]
\centering
\includegraphics[width=\columnwidth]{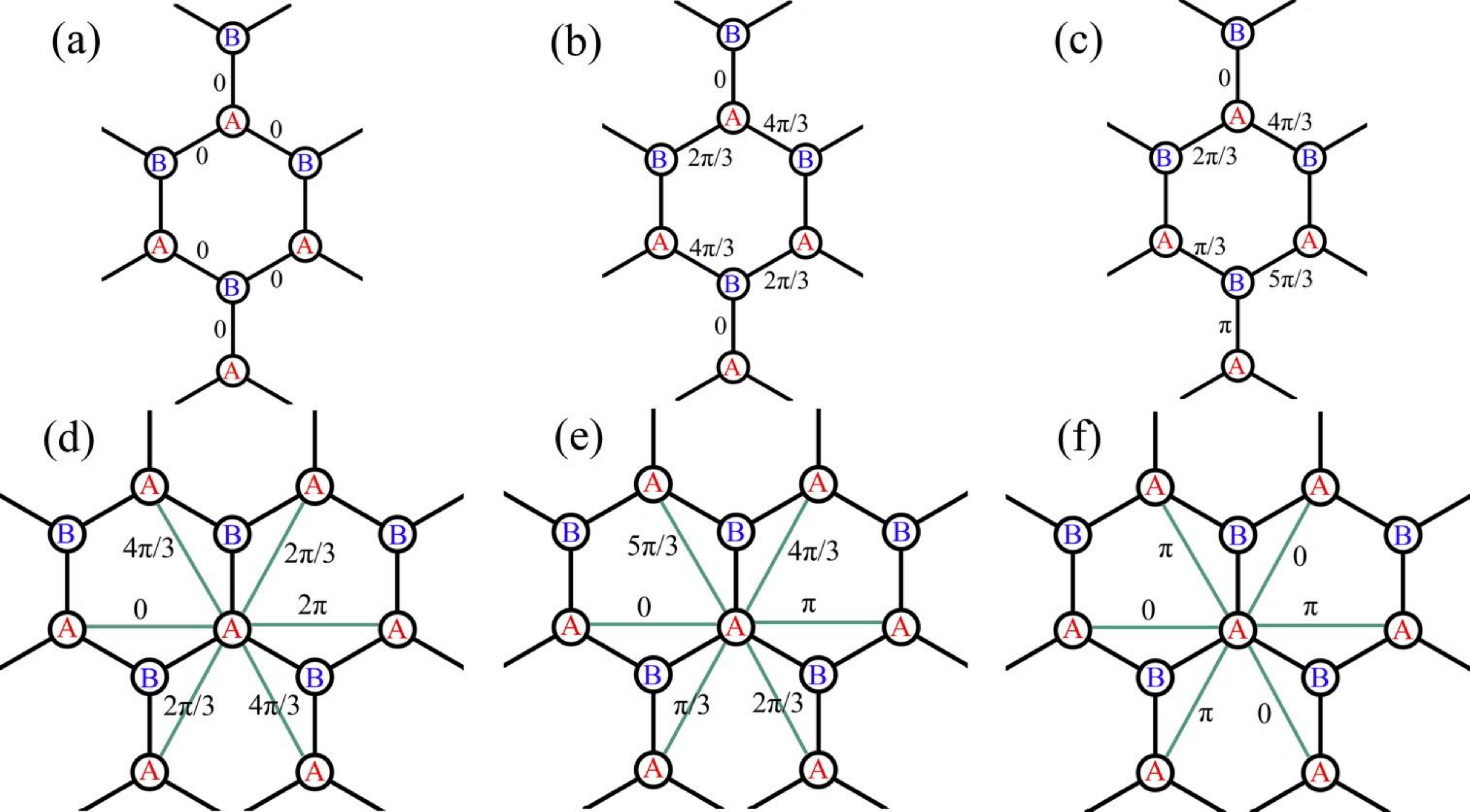}
\caption{(Color online) Phases of the considered pairing channels along the corresponding directions on the honeycomb lattice: (a) NN extended $s$-wave, (b) NN $d+id$-wave, (c) NN $p+ip$-wave, (d) NNN $d+id$-wave,
(e) NNN $p+ip$-wave and (f) NNN $f$-wave. \label{PairingChannels} }
\end{figure}

In order to probe for superconducting instabilities,
we examine the system's susceptibility towards various previously proposed pairing channels for this model. In particular, we consider the nearest-neighbor (NN)
extended $s$-wave, $d+id$-wave and $p+ip$-wave pairing correlations,  and consider also next-nearest-neighbor (NNN)
$d+id$-wave, $p+ip$-wave and $f$-wave pairings. In real space, these different pairing channels are
given in terms of appropriate form factors,
\begin{eqnarray}
f_{\text{NN},es} ({{\bm \delta}_l})&=&1,
\nonumber \\
f_{\text{NN},d+id} ({{\bm \delta}_l})&=&e^{i(l-1)\frac {2\pi}{3}},
\nonumber \\
f_{\text{NN},p+ip} ({{\bm \delta}_l})&=&e^{i(l-1)\frac {2\pi}{3}+\epsilon_s i \pi},
\nonumber \\
f_{\text{NNN}, d+id} ({{\bm \delta}'_l})&=&e^{i(l-1)\: \frac {2\pi}{3}},
\nonumber \\
f_{\text{NNN}, p+ip} ({{\bm \delta}'_l})&=&e^{i(l-1)\frac {\pi}{3}},
\nonumber \\
f_{\text{NNN}, f} ({{\bm \delta}'_l})&=&e^{i\frac{1+(-1)^l}{2}\pi}.
\label{formfactor}
\end{eqnarray}
where
the vectors ${\bm \delta}_l, l=1,2,3$  (${\bm \delta}'_l, l=1,2,...,6$) denote the different NN (NNN) lattice directions from a given lattice site, and  $\epsilon_s=0$ $(1)$ for sites on the A (B) sublattice. Figure~\ref{PairingChannels} shows these various form factors explicitly. In the spin sector, the $s$- and $d$-waves
are singlet states, while  $p$- and $f$-waves are triplet
states. The corresponding local pairing operators are  thus given as
\begin{equation}
\Delta_{\alpha\,{\bf i}}=
\frac{1}{\sqrt {N_\alpha}}\sum_{l} f_{\alpha} ({\bm \delta}^{(\prime)}_l)
(c^{} _{{\bf i}\uparrow}c_{{\bf i}+{\bm \delta}^{(\prime)}_l \downarrow}
\pm c^{} _{{\bf i}\downarrow}c_{{\bf i}+{\bm \delta}^{(\prime)}_l \uparrow}),
\end{equation}
where $+$ ($-$) for triplet (singlet) pairing, and
$N_\alpha$ are the corresponding normalization factors, with $N_\alpha=3$ ($N_\alpha=6$) for the NN (NNN) channels.

\begin{figure}[t]
\centering
\includegraphics[width=\columnwidth,angle=0,clip]{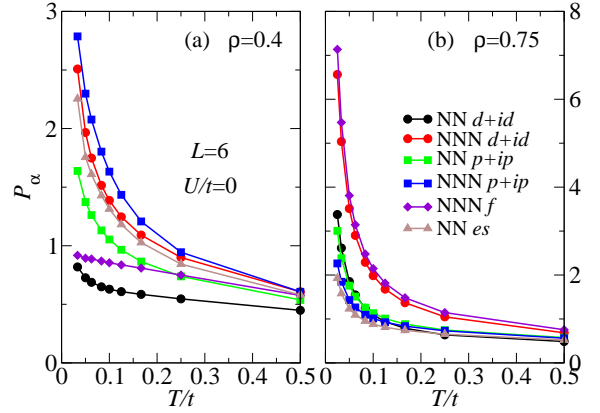}
\caption{(Color online) Temperature dependence of the various pairing susceptibilities $P_{\alpha}$ for the noninteracting system ($U=0$), as obtained on an $L=6$ system for (a)  $\rho=0.4$ and (b) $\rho=0.75$. \label{pairing_n6_U0_T} }
\end{figure}

Within the QMC simulations, we can directly access the temperature dependence of the  pairing susceptibilities for the various channels,
\begin{eqnarray}
P_{\alpha} &=&  \frac{1}{N_s}\sum_{{\bf i, \, j}} \int_0^\beta \,\, d\tau \,
\langle \Delta^\dagger_{\alpha\,{\bf i}}(\tau)
\Delta^{\phantom{\dagger}}_{\alpha\,{\bf j}}(0) \rangle,
\label{pairsusc}
\end{eqnarray}
where
$\Delta^\dagger_{\alpha\, {\bf i}}(\tau) =
e^{\tau H} \Delta^\dagger_{\alpha\, {\bf i}}(0) e^{-\tau H}$.
These pairing susceptibilities are however strongly
affected by the enhanced response of the free system at $U=0$.
This behavior is illustrated in Fig.~\ref{pairing_n6_U0_T}, which shows the different susceptibilities
$P_\alpha$ as functions of $T$ on the $L=6$ lattice, for both $\rho=0.4$ and $\rho=0.75$ in the noninteracting limit $U=0$. While there is
no superconducting ground states in the noninteracting case,
the apparent divergence of the $P_\alpha$ upon lowering $T$  provides a background to the
susceptibility measurements in the interacting case, in particular in the low-coupling regime that we can access in the FT-DQMC simulations.
We thus require to
examine the various pairing channels  based on the effective pairing interaction vertex~\cite{Scalapino89}.
In order to extract the corresponding effective pairing susceptibilities, we   compute in  FT-DQMC also the bare pairing
contributions $\tilde{P}_{\alpha}$,  for which two-particle terms
$\langle \,\, c^{\dagger}_{{\bf i}\, \downarrow}(\tau)
\,c_{{\bf j}\,\downarrow}^{\phantom{\dagger}}(0)
\,c^{\dagger}_{{\bf {k}}\, \uparrow}(\tau)
\,c^{}_{{\bf {l}}\,\uparrow}(0) \,\, \rangle$ that
appear  in evaluating the $P_{\alpha}$ in Eq.~\ref{pairsusc}
are  replaced by the decoupled contributions
$\langle \,\, c^{\dagger}_{{\bf i}\, \downarrow}(\tau)
\,c_{{\bf j}\,\downarrow}^{\phantom{\dagger}}(0)\,\,\rangle
\langle \,\,\,c^{\dagger}_{{\bf {k}}\, \uparrow}(\tau)
\,c^{}_{{\bf {l}}\,\uparrow}(0) \,\, \rangle$.
The effective pairing susceptibilities are  then given as $P^{\text{eff}}_{\alpha}=P_{\alpha}-\tilde{P}_{\alpha}$, and where  a
positive (negative) value of $P^{\text{eff}}_\alpha$ signals an enhanced (suppressed)  tendency towards pairing in
the corresponding channel. By definition, for the noninteracting case, the $P^{\text{eff}}_\alpha$ vanishes.

\begin{figure}[t]
\centering
\includegraphics[width=\columnwidth,angle=0,clip]{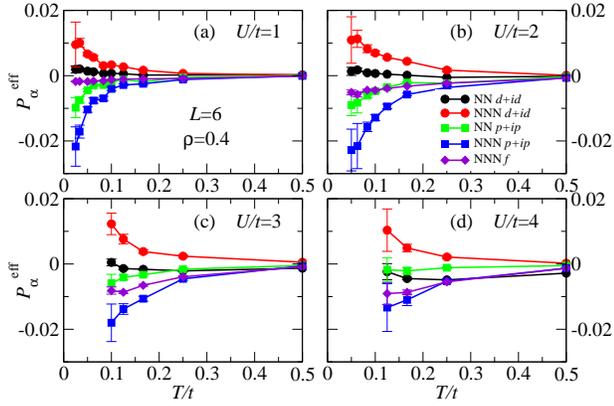}
\vspace{-0.5cm}
\caption{(Color online) Temperature dependence of the effective pairing susceptibilities $P^{\text{eff}}_\alpha$ at density $\rho=0.4$ on the $L=6$ lattice for (a) $U/t=1$, (b) $U/t=2$,
(c) $U/t=3$ and (d) $U/t=4$. \label{pairing_n6_rho04_T} }
\end{figure}

We now turn to examine the interacting system, and begin with the case of an electron density of $\rho=0.4$, i.e., well below the VHS filling.  First, we consider the results obtained for the $L=6$ lattice with 72 sites.
At this density the
sign problem is sufficiently moderate, and we can obtain the $P^{\text{eff}}_\alpha$ up to $U/t=4$, as shown in Fig.~\ref{pairing_n6_rho04_T} (a) to (d) for $U/t=1$ to $U/t=4$, respectively. These results for the $L=6$ lattice
exhibit that  consistently both   the NN and NNN $d+id$-wave
pairing susceptibilities are enhanced upon lowering $T$,  for all the considered interaction strengths (We also measured the extended $s$-wave channel susceptibility, but it is rather strongly suppressed in all the interacting cases that we considered and we thus do not include it in Fig.~\ref{pairing_n6_rho04_T}  or  any of the figures below).

\begin{figure}[t]
\centering
\includegraphics[width=\columnwidth,angle=0,clip]{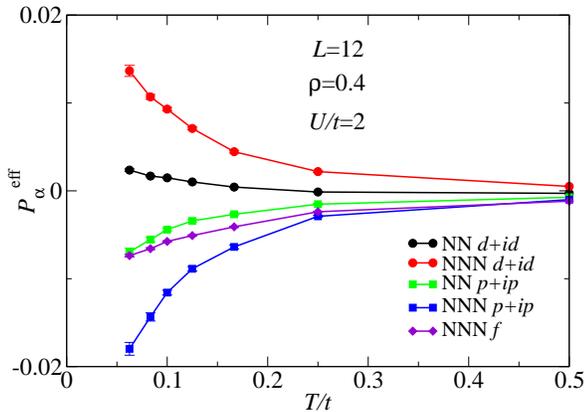}
\vspace{-0.5cm}
\caption{(Color online) Temperature dependence of the effective pairing susceptibilities $P^{\text{eff}}_\alpha$  at density $\rho=0.4$ on the $L=12$ lattice for  $U/t=2$. \label{pairing_n12_rho04_T} }
\end{figure}

To assess the stability of this result with respect to  finite size effects, we also performed  simulations on the $L=12$
system with 288 sites,  i.e.,  four times larger than the $L=6$ lattice.
Since in Fig.~\ref{pairing_n6_rho04_T} we find the prevailing pairing channel does not depend on the interaction strengths at $\rho=0.4$, we concentrate  in Fig.~\ref{pairing_n12_rho04_T} to the case of $U/t=2$ for  the $L=12$ lattice.
For $\rho=0.4$, the results on the $L=12$ lattice are in accord  with the findings on the $L=6$
lattice, and we conclude that  $d+id$-wave pairing forms the dominant pairing channel in this doping regime.
This is in good accord with various previous findings, as mentioned in the introduction.

We next perform a similar investigation for the VHS filling, $\rho=0.75$.
Due to the sign problem,
we are  in this case limited to weaker interactions, and consider explicitly here the cases of $U/t=1$ and $U/t=2$.
\begin{figure}[t]
\centering
\includegraphics[width=\columnwidth,angle=0,clip]{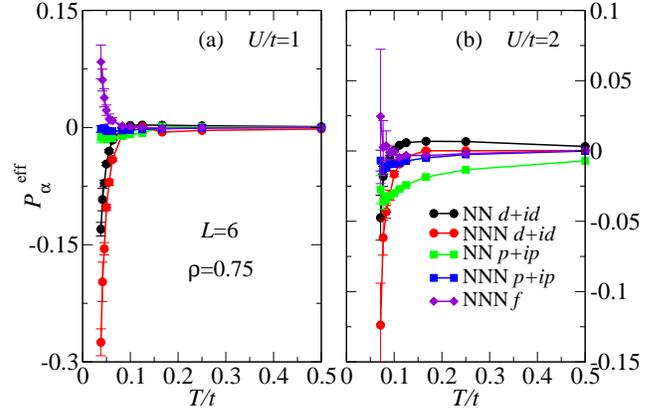}
\vspace{-0.5cm}
\caption{(Color online) Temperature dependence of the effective pairing susceptibilities at  the VHS filling ($\rho=0.75$) on the $L=6$ lattice for (a) $U/t=1$ and (b) $U/t=2$. \label{pairing_n6_rho075_T} }
\end{figure}
\begin{figure}[t]
\centering
\includegraphics[width=\columnwidth,angle=0,clip]{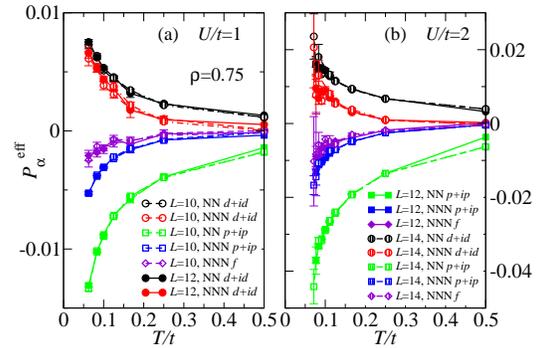}
\vspace{-0.5cm}
\caption{(Color online) Temperature dependence of the effective pairing susceptibilities  at  the VHS filling ($\rho=0.75$)  on the $L=10,12$ and $14$ lattices for (a) $U/t=1$ and (b) $U/t=2$.  \label{pairing_n12_rho075_T} }
\end{figure}
In contrast to the case of $\rho=0.4$, we observe strong finite-size effects  at the VHS filling, even regarding the leading low-temperature effective pairing susceptibility: As shown
in Fig.~\ref{pairing_n6_rho075_T}(a), for a
weak coupling of $U/t=1$ on the $L=6$ lattice, upon lowering the temperature, the effective
pairing susceptibility in the NNN $f$-wave channel gets strongly enhanced, while all other channels get suppressed, which suggests
 $f$-wave pairing  to dominate at  the VHS in the weak coupling region. If the interaction strength is increased
to $U/t=2$ in Fig.~\ref{pairing_n6_rho075_T}(b), the dominant pairing on the $L=6$ system  still appears in the  $f$-wave channel, however,  the
error bars are larger, due to a more severe sign problem.
Considering  the larger system sizes $L=10,12$ and $14$ at the VHS filling, shown in Fig.~\ref{pairing_n12_rho075_T}, we instead find --  consistently among these larger system sizes --  that the
dominant pairing channel switches from the  $f$-wave observed on the $L=6$ system to the NN and NNN $d+id$-wave pairings when the lattice size is increased.
The reason for this behavior may be the fact that on these larger lattice sizes, we resolve a more narrow grid of  momenta within the Brillouin zone, thus better resolving  the effective
interactions near the momenta corresponding to the VHS in the  density of states (DOS) -- which is most important at the VHS filling.

Another reason for this  size dependence may be that due to the enhanced DOS at the VHS filling, other electronic instabilities  compete with superconductivity. Indeed, based on a recent mean-field theory~\cite{Li12} and fRG calculations~\cite{wang12}, a particular interesting chiral SDW state was argued to form the leading magnetic instability of the Hubbard model at the VHS filling. In the following section, we  examine this scenario based on FT-DQMC simulations.

\section{Magnetic Correlations}
The chiral SDW state considered in Refs.~\onlinecite{Li12,wang12}
is characterized by the three independent nesting vectors ${\bf Q}_i$, $i=1,2,3$ of the free-system's Fermi surface at the VHS filling, which (folded back to the first Brillouin zone) correspond to the three independent $M$ points at the centers of
the Brillouin zone edges.  In terms of the reciprocal lattice vectors ${\bf b}_1$ and ${\bf b}_2$, these are
${\bf Q}_1=\frac{1}{2}{\bf b}_1$, ${\bf Q}_2=\frac{1}{2}{\bf b}_2$, ${\bf Q}_3=\frac{1}{2}({\bf b}_1+{\bf b}_2)$.
For lattice sites on the $A$ and $B$ sublattices within a unit cell centered at position $\bf R$,
the mean-field expectation values of the local spin operator in the chiral SDW state are proportional (up to a global rotation in spin space) to the  local direction vectors
\begin{eqnarray}
\langle {\bf S}_{{\bf R},A} \rangle_{\mathrm{cSDW}}&= &\frac{1}{\sqrt{3}}(\hat{\bm z} e^{i{\bf Q}_3 \cdot {\bf R}} + \hat{\bm x} e^{i{\bf Q}_1 \cdot {\bf R}} + \hat{\bm y} e^{i{\bf Q}_2 \cdot {\bf R}}),\nonumber \\
\langle {\bf S}_{{\bf R},B} \rangle_{\mathrm{cSDW}} & =&\frac{1}{\sqrt{3}}(\hat{\bm z} e^{i{\bf Q}_3 \cdot {\bf R}} - \hat{\bm x} e^{i{\bf Q}_1 \cdot {\bf R}} - \hat{\bm y} e^{i{\bf Q}_2 \cdot {\bf R}}),
\end{eqnarray}
where the $\hat{\bm x}$, $\hat{\bm y}$ and $\hat{\bm z}$ are the three mutually orthogonal unit vectors in spin space~\cite{Li12}. This state exhibits four different spin directions $\hat{\bm x} + \hat{\bm y} + \hat{\bm z}$, $-\hat{\bm x} -\hat {\bm y} +\hat {\bm z}$, $\hat{\bm x} - \hat{\bm y} -\hat {\bm z}$ and $-\hat{\bm x} + \hat{\bm y} -\hat {\bm z}$,  the magnetic  unit cell thus contains eight lattice sites, and we require $L$ to be even in order to accommodate this spin structure within the finite rhombic clusters.
In order to probe for this chiral SDW within the FT-DQMC simulations, we monitor a corresponding structure factor
\begin{equation}
\label{Schiral}
{S}_\mathrm{cSDW} = \frac{1}{N_s}\langle (\sum_{{\bf R}} M_{\bf R})^{\dagger} (\sum_{{\bf R}} M_{\bf R}) \rangle
\end{equation}
in terms of the projections
$M_{\bf R}=M_{\bf R, A} + M_{\bf R, B}$, with
$M_{\bf R, A(B)}= {\bf S}_{{\bf R},A(B)} \cdot \langle {\bf S}_{{\bf R},A(B)} \rangle_\mathrm{cSDW}$
of the local spin operators onto the chiral SDW texture. Here, ${\bf S}_{{\bf R},A(B)}$ denote the local spin operator on the A (B) sublattice site within the unit cell at position ${\bf R}$;  for a lattice site at position ${\bf i}$ this is given
as
${\bf S}_{\bf i}=\frac{1}{2}\sum_{\alpha,\beta}c^\dagger_{{\bf i},\alpha} {{\bm \sigma}}_{\alpha,\beta} c_{{\bf i},\beta}$
in terms of fermionic operators
and the vector  ${\bm \sigma}$ of Pauli matrices.
In the following, we consider for comparison also the corresponding antiferromagnetic structure factor $S_\mathrm{AF}$ for the antiferromagnetic N\'eel state, which is defined similarly to ${S}_\mathrm{cSDW}$, but with (up to a global spin rotation)  $\langle {\bf S}_{{\bf R},A} \rangle_{\mathrm{AF}}=\hat{\bm z}$, and
$\langle {\bf S}_{{\bf R},B} \rangle_{\mathrm{AF}}=-\hat{\bm z}$, respectively. The antiferromagnetic N\'eel state is well known to emerge in the half-filled system for sufficiently strong interactions.
However, here we first focus on the behavior of the chiral SDW structure factor ${S}_\mathrm{cSDW}$, considering the two specific electronic densities $\rho=0.75$ and $\rho=0.4$  as above.

\begin{figure}[t]
\centering
\includegraphics[width=\columnwidth,angle=0,clip]{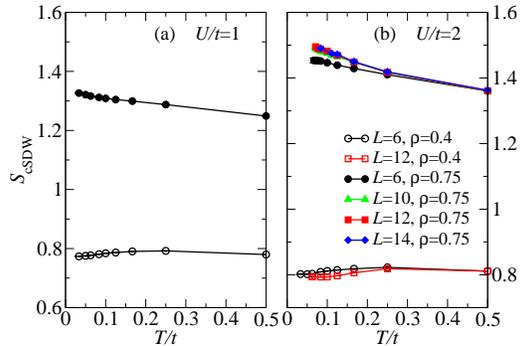}
\vspace{-0.5cm}
\caption{(Color online) Temperature dependence of the structure factor $S_\mathrm{cSDW}$ at $\rho=0.75$ and $\rho=0.4$, for (a) $U/t=1$
on the $L=6$ lattice and (b) $U/t=2$ for lattice sizes $L=6,10,12,14$. \label{Schiral_n_U_rho_T} }
\end{figure}

\begin{figure}[t]
\centering
\includegraphics[width=\columnwidth,angle=0,clip]{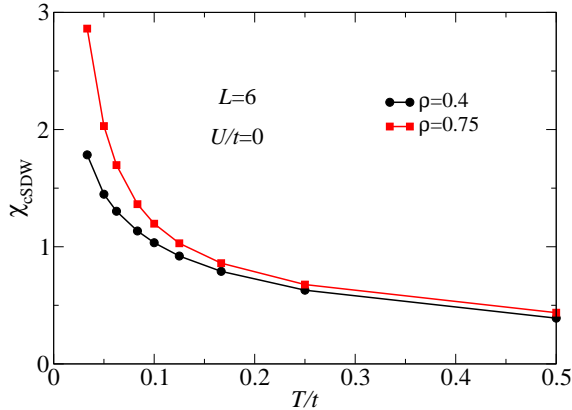}
\vspace{-0.5cm}
\caption{(Color online) Temperature dependence of  $\chi_{\mathrm{cSDW}}^{}$ for $\rho=0.4$ and $\rho=0.75$,
at  $U/t=0$  on the $L=6$ lattice. \label{Susp_n6_U0_rho_T} }
\end{figure}

In Fig.~\ref{Schiral_n_U_rho_T}, we show the FT-DQMC results for  $S_\mathrm{cSDW}$
as functions of $T$ for the two densities $\rho=0.75$ and $\rho=0.4$ at both $U/t=1$ and $U/t=2$ on the $L=6$
lattice. For $U/t=2$ we also performed  simulations on the $L=12$ lattice
(as well as on the $L=10$ and $14$ system for $\rho=0.75$)
in order to
examine finite-size effect in
$S_\mathrm{cSDW}$.
We find that upon lowering the temperature, $S_\mathrm{cSDW}$
increases at $\rho=0.75$, whereas it does not significantly increase, but is even weakly  suppressed at $\rho=0.4$.
We furthermore observe a mild increase of $S_\mathrm{cSDW}$ with system size $L$ at the VHS filling $\rho=0.75$.
Since the
corresponding magnetic instabilities can occur only at $T/t=0$ (due to the SU(2) symmetry of the Hamiltonian $H$), these results suggest that the chiral SDW
order, while possibly relevant at
$\rho=0.75$, is not favored at $\rho=0.4$.
A similar picture also emerges from analyzing   the
corresponding chiral SDW susceptibility
\begin{equation}
\label{chichiral}
\chi_{\mathrm{cSDW}}^{} =\frac{1}{N_s}\int_0^\beta \,\, d\tau \langle (\sum_{{\bf R}} M_{\bf R}(\tau))^{\dagger} (\sum_{{\bf R}} M_{\bf R}(0)) \rangle,
\end{equation}
where
$M_{{\bf R}}^{\dagger}(\tau) = e^{\tau H} M_{{\bf R}}^{\dagger}(0) e^{-\tau H}$.
Here, we need to again account for the enhanced response of the free system at $U=0$. This is shown in Fig.~\ref{Susp_n6_U0_rho_T}: for both densities, $\chi_{\mathrm{cSDW}}^{}$ at $U=0$ exhibits an apparent divergence upon lowering the temperature.
Similarly to the case of the pairing
susceptibilities, we thus examine  the corresponding effective chiral SDW susceptibility, which is
obtained as
$\chi_{\mathrm{cSDW}}^{\text{eff}}=\chi_{\mathrm{cSDW}}^{}-\tilde{\chi}_{\mathrm{cSDW}}^{}$, where
$\tilde{\chi}_{\mathrm{cSDW}}^{}$ denotes the  bare chiral SDW susceptibility. {This
procedure is similar to the antiferromagnetic case considered in Ref.~\onlinecite{Scalapino89}.}

\begin{figure}[t]
\centering
\includegraphics[width=\columnwidth,angle=0,clip]{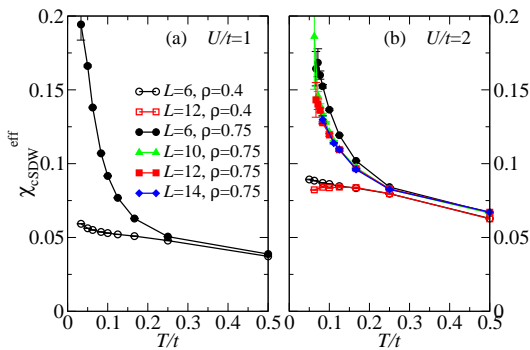}
\vspace{-0.5cm}
\caption{(Color online) Temperature dependence of   $\chi_{\mathrm{cSDW}}^{\text{eff}}$  at $\rho=0.4$ and $\rho=0.75$,
for (a) $U/t=1$ on a $L=6$  lattice, and (b) $U/t=2$ on $L=6,10,12$ and $14$ lattices. \label{Susp_chiral_n_U_rho_T} }
\end{figure}

As shown in Fig.~\ref{Susp_chiral_n_U_rho_T} for
$U/t=1$ and $U/t=2$, the effective susceptibility $\chi_{\mathrm{cSDW}}^{\text{eff}}$ at $\rho=0.75$ strongly increases
in the low-$T$ region
for all system sizes,
while $\chi_{{\mathrm{cSDW}}}^{\text{eff}}$ at $\rho=0.4$ does not show a similarly strong enhancement, whereas for $U/t=2$ it is even weakly suppressed at low $T$ for the
larger system size.
Unfortunately, the sign-problem does not allow us to perform
low-temperature simulations
on larger system sizes in order to perform a throughout finite-size scaling analysis of, e.g.,  $S_\mathrm{cSDW}$ at low temperatures,
which would be required in order to assess, if a chiral SDW ground state exists in the thermodynamic limit. Note that this case is different from the case of  pairing instabilities, which may in principle set in  at a {\it finite} (but still small) low-temperature scale.
Nevertheless, our findings provide indication that at the VHS filling the system may exhibit an
instability to the chiral SDW order, whereas
away from the VHS filling, this instability is eventually suppressed.

\begin{figure}[t]
\centering
\includegraphics[width=\columnwidth,angle=0,clip]{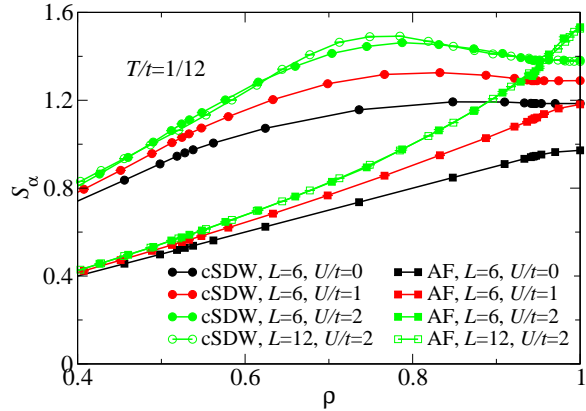}
\vspace{-0.5cm}
\caption{(Color online) Structure factors $S_\mathrm{cSDW}$ and $S_\mathrm{AF}$ as functions of $\rho$ for several values of $U/t$, for $T/t=1/12$ on the $L=6$ lattice. For $U/t=2$,  data for the $L=12$ lattice is also shown. \label{S_B12_n_U_rho} }
\end{figure}

To  investigate further how the chiral SDW order behaves at and beyond  the  VHS filling, we next fix an accessible, low
temperature $T/t =1/12$ and monitor how $S_\mathrm{cSDW}$ and $\chi_{\mathrm{cSDW}}^{\text{eff}}$  vary with the electronic density $\rho$. For this purpose,
Fig.~\ref{S_B12_n_U_rho} shows $S_\mathrm{cSDW}$  as a function of $\rho$ for different values of $U/t$.   These
results  indicate that upon increasing  $U/t$, a peak in $S_\mathrm{cSDW}$  gradually builds up near the VHS filling, such that the chiral SDW is indeed most pronounced at the VHS filling. This observation complies to the fact that the three characteristic momentum vectors ${\bf Q}_i$, $i=1,2,3$ of the chiral SDW state form the nesting vectors of the Fermi surface at the noninteracting system at the VHS filling.
For comparison, we also show in this figure the antiferromagnetic structure factor $S_\mathrm{AF}$,
which in contrast to $S_\mathrm{cSDW}$  displays a monotonic increase with increasing electron density.
At half-filling, $\rho=1$, the Hubbard model on the honeycomb lattice is well known to harbor a quantum
phase transition to an insulating antiferromagnetic  phase for $U/t>3.76$\cite{Sorella12}.
While in Fig.~\ref{S_B12_n_U_rho}, we remain below this critical value of $U$,  the antiferromagnetic  correlations
already display a clear  tendency to grow with increasing $U$. Furthermore,
at $U/t=2$, the antiferromagnetic  structure factor  exceeds the chiral SDW structure factor at (and close to) half-filling,
while upon doping further below half-filling, towards the VHS filling,  the chiral SDW correlations become more dominant.

\begin{figure}[t]
\centering
\includegraphics[width=\columnwidth,angle=0,clip]{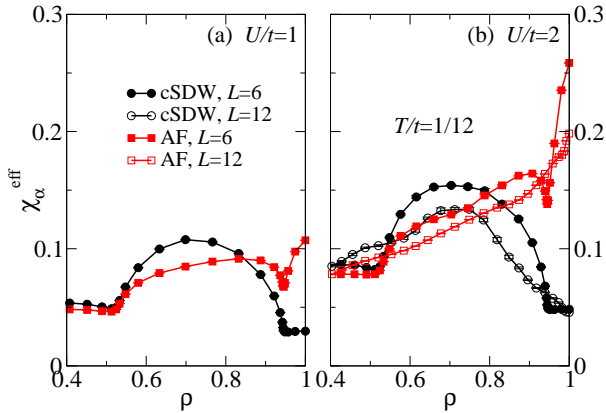}
\vspace{-0.5cm}
\caption{(Color online) Effective susceptibilities $\chi_\mathrm{cSDW}^{\text{eff}}$ and $\chi_\mathrm{AF}^{\text{eff}}$ as  functions of $\rho$ at $T/t=1/12$, for (a) $U/t=1$
on the $L=6$ lattice, and (b) $U/t=2$
on both $L=6$ and $L=12$ lattices.\label{Susp_B12_n_U_rho} }
\end{figure}

We observe a similar enhancement in the chiral SDW response near the VHS filling  also for the effective susceptibility
$\chi_{\mathrm{cSDW}}^{\text{eff}}$, cf. Fig.~\ref{Susp_B12_n_U_rho}, strengthening the above interpretation of the structure factor data.
Note that in Fig.~\ref{Susp_B12_n_U_rho}, the $L=6$ data exhibits two kinks around $\rho\approx 0.5$ and $\rho\approx 0.95$. These  appear to be due to finite-size effects -- compare to the data for the $L=12$   lattice, where both kinks are absent. Such peculiar finite-size effects can in fact also be observed in a plot of the electronic density as a function of  the chemical potential $\mu$ in Fig.~\ref{rho_U2_B12_n_mu}:
\begin{figure}[t]
\centering
\includegraphics[width=\columnwidth,angle=0,clip]{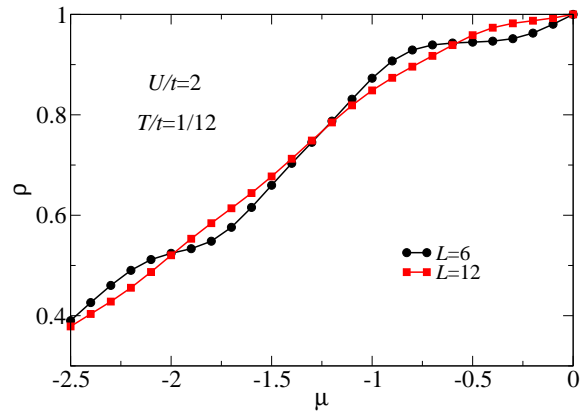}
\vspace{-0.5cm}
\caption{(Color online) Electronic density $\rho$ as a function of the chemical potential $\mu$ on the $L=6$ and $L=12$ lattices, for $U/t=2$ and $T/t=1/12$.\label{rho_U2_B12_n_mu} }
\end{figure}
on the $L=6$  lattice, the density as a function of $\mu$ shows two plateaus near $\rho=0.5$ and $\rho=0.95$,
whereas on the larger lattice, those plateaus have disappeared. We consider these finite-size plateaus to be
the reason also for the two kinks seen in Fig.~\ref{Susp_B12_n_U_rho} for the $L=6$  lattice.
For the $L=12$ lattice the density plateaus are absent, and
$\chi_{\mathrm{cSDW}}^{\text{eff}}$ decreases steadily upon doping away from the VHS filling, again suggesting that  the chiral SDW instability  is important  when the filling is at (and maybe also close to) the VHS value. For comparison, the effective antiferromagnetic susceptibility $\chi_\mathrm{AF}^{\text{eff}}$ is also shown in Fig.~\ref{Susp_B12_n_U_rho}
(where $\chi_\mathrm{AF}^{\text{eff}}$ is defined similarly as the effective susceptibility for the chiral SDW case). While on the $L=6$ system,
this quantity  shows similar finite-size anomalies as  the effective chiral SDW susceptibility
$\chi_{\mathrm{cSDW}}^{\text{eff}}$, on the $L=12$ system it instead shows a monotonic decease when doping away from half-filling, as anticipated from the behavior of the antiferromagnetic structure factor.

\section{Summary}

To conclude, we used finite-temperature determinantal quantum Monte Carlo simulations  to examine  the electronic  pairing channels  and magnetic instabilities of doped graphene within the Hubbard model description.
Due to the  sign problem,  we restricted to the weak coupling regime at the VHS filling, while at lower fillings beyond the VHS,
we  also accessed  the weak to intermediate coupling regime.
In both cases, we find NN and NNN $d+id$-wave pairing
as  the dominant pairing channels on the larger system sizes. However, at the VHS filling, we observed strong finite-size effects in the dominant pairing symmetry. This may be taken as indication, that at this filling, due to the logarithmically diverging density of state and a nested Fermi surface also other electronic instabilities may be relevant. In fact, we observe from measuring appropriate structure factors and magnetic susceptibilities that a previously proposed chiral spin density wave state shows a robust enhancement near the VHS filling, but weakens quickly upon doping away from the VHS point. This
 is  in  accord with the  result in Ref.~\onlinecite{Lothman17}, which suggests on the mean-field level that upon doping away from a DOS peak, instabilities within the  particle-particle channel (superconducting orders) survive decisively further than those in the particle-hole channel (magnetic or charge
orders). We note that a previous study of the Hubbard model on the triangular lattice reported a related result in terms of ferromagnetism and $f$-wave pairing.~\cite{Su08}
For the future, it will be interesting to extend also  dynamical cluster approximation studies to consider the competition among the superconducting and magnetic instabilities of the doped honeycomb lattice Hubbard model.

\section*{ACKNOWLEDGMENTSS}

We thank C. Honerkamp and Z. Y. Meng for useful discussions. This work is supported by the Deutsche Forschungsgemeinschaft (DFG) through the grants FOR 1807 and RTG 1995.  T. Y.  is also supported by the National Natural Science Foundation of China (NSFC Grants No. 11504067). Furthermore, we thank the IT Center at RWTH Aachen University and the JSC J\"ulich for access to computing time through JARA-HPC.

\end{document}